\documentclass[aps,prd,showpacs,amsmath,amssymb]{revtex4}
\usepackage{epsfig}

\usepackage{graphicx}
\usepackage{dcolumn}
\usepackage{bm}

\def\be{\begin{equation}}
\def\ee{\end{equation}}
\def\te{\end{equation}}
\def\bea{\begin{eqnarray}}

\def\tea{\end{eqnarray}}
\def\SIm{\mathrm{Im}}

\begin{document}

\title{Vortex Formation in Two-Dimensional Bose Gas}
\author{Esteban Calzetta}
\email{calzetta@df.uba.ar} \affiliation{IFIBA (CONICET) and Departamento de
Fisica, Facultad de Ciencias Exactas y Naturales, Universidad de
Buenos Aires- Ciudad Universitaria, 1428 Buenos Aires, Argentina}
\author{Kwan-yuet Ho}
\email{kwyho@umd.edu} \affiliation{Institute of Physical
Sciences and Technology and Department of Physics, University of
Maryland, College Park, Maryland 20742-4111, USA}
\author{B. L. Hu} \email{blhu@umd.edu}
\affiliation{Joint Quantum Institute and Maryland Center for
Fundamental Physics, University of Maryland, College Park, Maryland
20742-4111, USA}
\date{October 21, 2009}

\begin{abstract}
We discuss the stability of a homogeneous two-dimensional
Bose gas at finite temperature against
formation of isolated vortices. We consider a patch of several
healing lengths in size and compute its free energy using the
Euclidean formalism. Since we deal with an open system, which is able to
exchange particles and angular momentum with the rest of the
condensate, we use the symmetry-breaking (as opposed to the particle
number conserving) formalism, and include configurations with all
values of angular momenta in the partition function. At finite
temperature, there appear sphaleron configurations associated to
isolated vortices. The contribution from these configurations to the
free energy is computed in the dilute gas approximation. We show that
the Euclidean action of linearized perturbations of a vortex is not
positive definite. As a consequence the free energy of the 2D Bose gas
acquires an imaginary part. This signals the instability of the gas.
This instability may be identified with the Berezinskii, Kosterlitz
and Thouless (BKT) transition.
\end{abstract}



\maketitle

\section{Introduction} \label{sec_intro}

Below a certain temperature in a three-dimensional Bosonic system,
long range order is established in an ordered phase called the
Bose-Einstein condensate (BEC). In two-dimensional (2D) Bose gas,
there does not exist an ordered state since its existence means that
the correlation of its fluctuations is logarithmically divergent, as proven by Mermin,
 Wagner \cite{MerWag66}, Hohenberg \cite{Hoh67}
and Coleman \cite{Col73}. However, it has been shown by Berezinskii \cite{Ber71}, Kosterlitz
and Thouless \cite{KosTho73} (BKT) that there exists a superfluid
phase with a quasi-long-range order below a certain temperature. The
superfluid phase has only the bounded vortex pairs but above the BKT
temperature single vortices proliferate as this is the more stable
configuration \cite{KosTho73}.

Many theoretical studies on BKT transition in 2D Bose gas, including
the original theory of BKT, are based on equilibrium thermodynamics of
many-body systems \cite{KAPS02, HowLeC10, GraWet95, GerWet01,
FloWet09a}.  There have been studies of 2D vortex dynamics, viewed as
massive charged particles in relativistic two-dimensional
electrodynamical systems \cite{Pop73, Pop87_Ch8}. This analogy has been
applied to the study of vortex dynamics of 2D superfluid via a
Fokker-Planck equation \cite{ChuWil00} and using field theoretical
approaches \cite{AroAue08, LiAuAr08, WaDuMa10}. There were also
numerical studies based on the Gross-Pitaevskii equation \cite{Gro63,
GinPit58} on the lifetime of spontaneous decay of a pancake-shaped
condensate with a vortex \cite{PLEB99}. 
There were also studies around the critical region of BKT transition with Monte Carlo simulations \cite{PrRuSv01, ProSvi02} with the local density approximation.
With the use
of projected Gross-Pitaevskii equation (PGPE) \cite{DaMoBu02}, the thermal activation of vortex
pairs in the presence of a harmonic trap \cite{HutBla06, SimBla06, SchHut07} and 
its emergence of superfluidity \cite{SiDaBl08} were studied, and the various consequences
of the improved mean-field Holzmann-Chevalier-Krauth (HCK) theory \cite{HoChWe08} 
of the 2D Bose gas \cite{BiBaBl09, BDSB09, BisBla09a, BisBla09b} were presented. The
non-equilibrium response of a 2D Bose gas is less understood. One       
encounters such a situation when the trap is suddenly turned off, as
is done in recent experiments \cite{HKCBD06, KrHaDa07, CRRHP09} described below.

Recently 2D quantum Bose gas has been experimentally realized by the
Dalibard group \cite{HKCBD06, KrHaDa07} by slicing a 3D
BEC into pieces of ``pancakes'' with 1D optical lattices, and by the
Phillips group \cite{CRRHP09} through trapping the atoms in a 3D
harmonic potential with a very large frequencies in one of the directions. In
both experiments, measurements on the gas are performed some time
after the confining potential is abruptly turned off. Dalibard
group's experiment showed that there are more isolated vortices
formed at higher temperatures. 
The Phillips group measured the density profile after 10 ms time of flight
, and identified different states of the gas.
In one regime the gas develops a bimodal distribution with only
thermal and quasi-condensate components without long range order, as different from a superfluid. For a sufficiently long time of flight, they observe a trimodal distribution with thermal, quasi-condensate and superfluid components indicative of a BKT transition

In this paper, we compute the free energy of a 2D Bose gas by means
of thermal field theory. We consider the action in the Madelung
representation (in terms of density and phase), and convert it to a
Euclidean action by a Wick rotation in time and in phase. The system
that we study is a patch of a size of several healing lengths within the
larger 2D gas. 
Because we are dealing with the homogeneous configuration, we
put no confining potential, i.e., $V(\mathbf{x})=0$. Since the vortices form at the center of the gas patch \cite{BiBaBl09} at the beginning where the density of the gas is effectively homogeneous, and the vortex core structure is very small compared to the size of the gas patch, we expect to reduce the physically more relevant inhomogeneous
situation to the homogeneous situation discussed here through a local density approximation. 
 Since particle number and angular momentum are not
conserved for this system, we do not constraint the former (unlike in the
particle number conserving formalism, see e.g., \cite{CalHu08_Sec13_3})
 and consider configurations with all values of angular
momentum. In particular, we consider configurations with different
numbers of vortices and the fluctuations around them.
Although these configurations are time-independent, they have finite euclidean action as a consequence of the compactification of the euclidean time axis, namely euclidean time is periodic with periodicity $\hbar\beta$. These time independent configurations with nonzero angular momentum play in our problem the same role as the usual sphaleron configurations in electroweak symmetry breaking \cite{Rub02}.
The contribution from these configurations to the free energy is computed
within the dilute gas approximation.

We find that the Euclidean action for  fluctuations around an
isolated vortex is not positive definite. In real time, this means an
instability of the isolated vortex, and we characterize the direction
of greatest instability in configuration space. In imaginary time the
fact that the Euclidean action is not positive definite means that
the partition function must be defined by an analytic continuation, whereby
the free energy becomes complex.  We calculate the imaginary part of
the free energy due to this instability. This is similar to the
argument of Langer who considered the decay of a metastable state due
to classical fluctuations \cite{Lan68, Lan69}, that of Coleman who
considered the quantum fluctuations around the spatially-separated
instantons \cite{Col77, CalCol77, Col85_Ch7}, and that of Affleck who
considered the decay of a quantum-statistical metastable state using
instantons \cite{AffDeL79, Aff81}. We find that the canonical 2D Bose
gas is indeed unstable at finite temperature, and the decay rate,
which is also the rate of vortex nucleation, increases with
temperature. For $T>T_{BKT}$, the gas evolves to a state of isolated
vortices.

The paper is organized as follows. In Sec. \ref{sec_GPtreatment} we introduce
the Gross-Pitaevskii treatment and write it in the Madelung representation.
In Sec. \ref{sec_euclaction} we obtain the Euclidean action by a Wick rotation and 
model the density profile of the gas with a vortex at the origin. In Sec. \ref{sec_linpert}
we introduce the linear perturbation about the configuration for each $q$. In Sec. \ref{sec_lifetime}
we outline the formalism of computing the lifetime of the gas and obtain
the BKT transition temperature. In Sec. \ref{sec_imF} we use Bohr-Sommerfeld quantization to show that the effective energy is complex, indicating the instability of the 2D Bose gas.
We end with conclusions in Sec. \ref{sec_summary}.

\section{Model} \label{sec_GPtreatment}

The dynamics of a two-dimensional (2D) Bosonic atomic system  with a
$\delta$-potential inter-atomic interaction is described by the
action \cite{Gro63,GinPit58}
\begin{equation}
S=\int\:dt\:d^2{\mathbf x}\:\left\{i\hbar\Psi^{\dagger}\frac{\partial \Psi}{\partial t}-H\right\}
\label{action} ,
\end{equation}
where $\Psi(\mathbf{x})$ and $\Psi^{\dag}(\mathbf{x})$ are respectively the annihilation and creation operators of an atom at point $\mathbf{x}$. The Hamiltonian is
\begin{equation}
H=\frac{\hbar^2}{2m}\nabla\Psi^{\dagger}\nabla\Psi +F\left[\Psi^{\dagger}\Psi\right]
\label{hamiltonian} ,
\end{equation}
and
\begin{equation}
F\left[\rho\right]=\left(V\left({\mathbf x}\right)-\mu\right)\rho +\frac12g\rho^2
\label{freeenergy} .
\end{equation}
where $g$ is the coupling constant due to the $\delta$-potential between the atoms. In the Madelung representation
\begin{equation}
\Psi =\sqrt{\rho}e^{i\theta}
\label{Madelung} ,
\end{equation}
the density of atoms in the lowest macroscopically occupied state $\rho$ and the phase $\theta$ are canonical to each
other, obeying the commutation relation \cite{Hal81, CaHuRe06}
\begin{equation}
\label{commutator_rhotheta} [\rho(\mathbf{x}), \varphi(\mathbf{x}')] = -i \delta(\mathbf{x}-\mathbf{x}') .
\end{equation}
With (\ref{Madelung}), the action (\ref{action}) is written as
\begin{equation}
S=\int\:dt\:d^2{\mathbf x}\:\left\{\hbar\theta\frac{\partial\rho }{\partial t}-H\right\} ,
\label{action2}
\end{equation}
where
\begin{equation}
H=\frac{\rho}{2m}\left(\nabla\hbar\theta\right)^2 +F_q\left[\rho\right] ,
\label{hamiltonian2}
\end{equation}
and
\begin{equation}
F_q\left[\rho\right]=F\left[\rho\right]+\frac{\hbar^2}{8m\rho}\left(\nabla\rho\right)^2 .
\label{quantumfree}
\end{equation}
The length scale that characterizes the local alteration of the gas density healing back to the mean-field density is given by the healing length, which is
\begin{equation}
\label{def_xi} \xi^2 = \frac{\hbar^2}{4 m \mu} ,
\end{equation}

Experimentally there is a harmonic trap to prepare the initial patch
of Bose gas in two-dimensions. At the time when the trap is turned
off, the Bose gas is still highly inhomogeneous. However, in recent
experiments, the vortex core structure is very small compared to the
patch of quasi-two-dimensional Bose gas. Take Phillips group's
experiment for example. Sodium atom is used and therefore $m \sim
3.8\times10^{-26}$ kg. And $\omega_{\bot} = 20$ Hz and $\omega_z = 1$
kHz \cite{CRRHP09}. It is known that $\mu = g \rho_0$, and $\rho_0
=\frac{4}{\lambda^2}$ \cite{PrRuSv01} and $\lambda =
\sqrt{\frac{2\pi\hbar^2}{mk_B T}}$ being the thermal de Broglie
wavelength. Near the transition point, $T \sim 100$ nK \cite{KrHaDa07,
  CRRHP09}, $\rho_0 \sim 3\times 10^{12}$ m${}^{-2}$. By $g =
\frac{\hbar^2}{m} \frac{a}{L_z}$ (where $a$ is the scattering length
and $L_z$ is the thickness of the gas) and the fact that for most
current experiments $\frac{a}{L_z} \sim \frac{1}{30}$ \cite{HadDal09},
$g \sim 9.7 \times 10^{-45}$ J m${}^2$. Then $\mu \sim 2.9 \times
10^{-32}$ J. Then from (\ref{def_xi}), $\xi^2 \sim 2.5 \times
10^{-12}$ m${}^2$. The area of the gas is given by the circle of the
TF radius, $A \sim \pi R_{\bot}^2 \sim
\frac{2\pi\mu}{m\omega_{\bot}^2} \sim 1.20 \times 10^{-8}$
m${}^2$. Hence $A >> \xi^2$, which means the vortex structure is very
small compared to the size of the gas. Hence, the experimental
situation can be recovered from our subsequent analysis through local
density approximation \cite{ProSvi02} that locally the gas is
effectively homogeneous at the center of the trap \cite{BiBaBl09}, which is best
described by $V(\mathbf{x})=0$. 

\section{Euclidean Action} \label{sec_euclaction}

To compute the partition function of such a system, we perform a Wick
rotation by writing $t = - i \tau$. To preserve the same canonical
relation between the density and the phase
(\ref{commutator_rhotheta}) and to keep the density real, the phase
has to be rotated accordingly by
\begin{equation}
\label{rotatephase} \chi = -i \hbar \theta .
\end{equation}
whence $\exp\left(i \frac{S}{\hbar}\right)$ becomes $\exp\left(-
\frac{\mathcal{S}}{\hbar}\right)$. The action in this Euclidean space
is given by
\begin{equation}
\mathcal{S}=\int\:d\tau\:d^2{\mathbf x}\:\left\{\chi \frac{\partial\rho }{\partial \tau}+\mathcal{H}\right\} ,
\label{action3}
\end{equation}
where the Hamiltonian density is
\begin{equation}
\mathcal{H}=-\frac{\rho}{2m}\left(\nabla\chi\right)^2 + F_q\left[\rho\right] .
\label{hamiltonian3}
\end{equation}

Let us introduce the following dimensionless variables
\begin{equation}
\label{def_dimensionless} \tau =\frac{\hbar}{\mu}s,\\
\mathbf{r}=\xi\mathbf{y},\\ \rho =\frac{\mu}{g} n ,
\end{equation}
and the Euclidean phase
\begin{equation}
\label{def_zeta} \chi = \hbar \zeta .
\end{equation}

With theses new variables the Euclidean action (\ref{action3})
becomes
\begin{equation}
\mathcal{S} = \frac{\hbar\mu\xi^2}{g} \int\:ds\:d^2{\mathbf y}\: \left\{\zeta\frac{\partial n }{\partial s}-2n\left(\nabla_y\zeta\right)^2 -n+\frac{n^2}2+\frac{\left(\nabla_y n\right)^2}{2 n}\right\} .
\label{expr_action}
\end{equation}

Because $\Psi$ is a single-valued function its value is unchanged upon having 
the phase $i \xi$ added by $2 \pi q$, for any integer $q$, 
to it does not change the
value of the field. As a result, for any integer $q$,
\begin{equation}
\label{def_vortex} \oint d\mathbf{l} \cdot \nabla_y \zeta = -2 \pi iq,
\end{equation}
where the line integral goes around a loop about a point. If the
vorticity $q$ is positive (negative) while the loop is small enough,
there is a vortex (an antivortex) at that point whereas $q=0$
indicates there is no vortex at that point. But if the loop of the
line integral is larger, $q$ is the sum of the vorticities of all
vortices inside the loop, while vortex and antivortex cancel each
other in the integration. The phase may have a curl-free part even if
there is a vortex. The simplest configuration representing a single 
vortex at the origin has $\zeta = -iq \varphi$. The Euclidean angular
momentum density of the system is given by
\begin{equation}
\label{def_l} l = \rho \frac{\partial \chi}{\partial \varphi} = -i\hbar q \rho ,
\end{equation}
which is proportional to $q$. The fact that the angular momentum
commutes with the Hamiltonian and is conserved implies the
conservation of vorticity in the whole system.

Assuming there is a vortex at the origin with density profile
$n_q(y)$, presumed to be rotationally invariant,the equation of
motion  is obtained by putting $\zeta = -iq \varphi$ into
(\ref{expr_action}):
\begin{equation} \label{eqn_n_q} \frac{1}{y}
\frac{d}{dy} \left( y \frac{d n_q}{d y} \right) - \frac{1}{2 n_q}
\left(\frac{d n_q}{dy}\right)^2 + \left(1 - \frac{2 q^2}{y^2}\right)
n_q - n_q^2 = 0 .
\end{equation}
For $q=0$, $n_q = 1$ exactly. In the general case, it is convenient
to introduce an ``Euclidean wave function of the condensate'' by
writing $n_q=\psi_q^2$ \cite{GinPit58, LifPit90_Ch30}. It then becomes

\be
\label{eqn_psi} \frac{1}{y} \frac{d}{dy} \left( y \frac{d \psi_q}{d y} \right)  + \frac12\left(1 - \frac{2 q^2}{y^2}\right) \psi_q - \frac12\psi_q^3 = 0 .
\end{equation}

The vortex solution interpolates between the no-vortex profile
$\psi_q=1$ for $y\mapsto\infty$ and the trivial solution $\psi_q=0$
for $y\mapsto 0$. Eq. (\ref{eqn_psi}) may be solved numerically (see
\cite{GinPit58, LifPit90_Ch30}). For large $y$, we may expand $\psi_q$ in
inverse powers of $y^2$:

\begin{equation}
\label{nq_largey} \psi_q = 1 - \frac{ q^2}{y^2} -\frac{\left[8q^2+q^4\right]}{2y^4} + O\left(\frac{1}{y^6}\right) .
\end{equation}
Likewise for the density:

\begin{equation}
\label{nq_largey2} n_q = 1 - \frac{ 2q^2}{y^2} -\frac{8q^2}{y^4} + O\left(\frac{1}{y^6}\right) .
\end{equation}

\begin{figure}[htp]
\centering
\includegraphics[scale=.8]{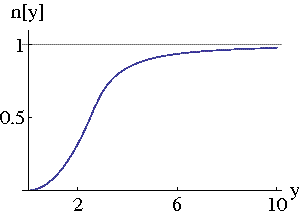}
\caption[g31] {The density profile for an isolated vortex at the
origin, as given by (\ref{nq_smally}) for $y<2.7$ and
(\ref{nq_largey}) for $y>2.7$.} \label{g31}
\end{figure}

For $y << 1$, the cubic term in (\ref{eqn_psi}) can be neglected, and
$\psi_q$ becomes a Bessel function \cite{BaChRe10}. For our purposes, it is enough to keep only the first (linear)
term in the Taylor expansion of $\psi_q $. The density profile is
then quadratic
\begin{equation}
\label{nq_smally} n_1 =0.08\:y^2.
\end{equation}
We shall adopt the approximation (\ref{nq_largey2}) for $y>2.7$ and
(\ref{nq_smally}) otherwise. The matching point and the constant in
(\ref{nq_smally}) are chosen so the approximated density profile is
smooth (see Fig. \ref{g31})

\section{Linear Perturbation} \label{sec_linpert}

Consider linear perturbations around a configuration of the 2D Bose
gas with a vortex at the origin:
\begin{equation}
\label{linpert} n = n_q (1+\delta),  \zeta = -iq \varphi + \zeta_1 ,
\end{equation}
where $\delta = \delta(y,\varphi,s)$ and $\zeta_1 = \zeta_1(y,\varphi,s)$ are functions of the radial and azimuthal coordinates $y$ and $\varphi$. Define the operator
\begin{equation}
\label{def_laplaciantilde} \tilde{\nabla}_y^2 = \frac{1}{y n_q} \frac{\partial}{\partial y} \left( y n_q \frac{\partial }{\partial y}\right) + \frac{1}{y^2} \frac{\partial^2 }{\partial \varphi^2} .
\end{equation}
Note that for $q=0$ (i.e., $n_q = 1$), $\tilde{\nabla}_y^2 =
\nabla_y^2$.  Putting the perturbation (\ref{linpert}) in the action
(\ref{expr_action}), it becomes
\begin{eqnarray}
\nonumber \mathcal{S} (q) \approx \mathcal{F}_0 (q) &+& \frac{\hbar \mu \xi^2}{g} \int_0^{\mu \beta} ds \int d^2 y \cdot n_q \delta \left(\frac{n_q}{2} - \frac{1}{2} \tilde{\nabla}_y^2 \right) \delta \\
\label{expr_action5} &+& \frac{\hbar \mu \xi^2}{g} \int_0^{\mu \beta} ds \int d^2 y \cdot \zeta_1 \left(2 n_q \tilde{\nabla}_y^2 \right) \zeta_1 \\
\nonumber &+&  \frac{\hbar \mu \xi^2}{g} \int_0^{\mu \beta} ds \int
d^2 y \cdot n_q \zeta_1 \left( \frac{\partial}{\partial s} - \frac{4i
q}{y^2} \frac{\partial}{\partial \varphi}  \right) \delta  ,
\end{eqnarray}
where
\begin{eqnarray}
\nonumber \mathcal{F}_0(q) &=& \frac{\hbar \mu \xi^2}{g}
\int_0^{\mu\beta} ds \int d^2 y \cdot \left[\frac{2 n_q q^2}{y^2} - n_q + \frac{n_q^2}{2} + \frac{1}{2 n_q} \left(\frac{d n_q}{d y}\right)^2 \right] \\
\label{def_F0} &=& - \frac{ \pi \hbar \xi^2 \mu^2 \beta}{g} \int dy
\cdot y n_q^2  ,
\end{eqnarray}
which is the equilibrium free energy, with the second equality owing
to (\ref{eqn_n_q}). Then the equations of motion are given by
\begin{eqnarray}
\label{lineqnmotion1} \frac{\partial \zeta_1}{\partial s} = n_q \delta + \frac{4i q}{y^2} \frac{\partial \zeta_1}{\partial \varphi} - \tilde{\nabla}_y^2 \delta  , \\
\label{lineqnmotion2} \frac{\partial \delta}{\partial s} = \frac{4i q}{y^2} \frac{\partial \delta}{\partial \varphi} - 4 \tilde{\nabla}_y^2 \zeta_1 .
\end{eqnarray}

The Fourier transform of the fluctuations can be defined as
\begin{eqnarray}
\label{def_fourierdelta} \delta (y, \varphi, s) = \sum_{j=-\infty}^{\infty} \delta_j(y, s) e^{i j \varphi}, \\
\label{def_fourierzeta1} \zeta_1 (y, \varphi, s) = \sum_{j=-\infty}^{\infty} \zeta_{1j}(y, s) e^{i j \varphi} .
\end{eqnarray}
If $\delta$ and $\zeta_1$ are real, then
\begin{equation}
\label{cplxconj} \delta_j^* = \delta_{-j}, \zeta_{1j}^* = \zeta_{1-j} .
\end{equation}
The representation (\ref{def_fourierzeta1}) assumes that
\begin{equation}
\label{curlless} \oint d\mathbf{l} \cdot \nabla \zeta_{1} = 0,
\end{equation}
which means the fluctuation does not change the total vorticity. The counterpart of $-\tilde{\nabla}_y^2$ in the Fourier representation is
\begin{equation}
\label{def_Ljq} L_{jq} = -\frac{1}{y n_q} \frac{\partial}{\partial y} \left(y n_q \frac{\partial}{\partial y}\right) + \frac{j^2}{y^2}.
\end{equation}
The action (\ref{expr_action5}) becomes
\begin{eqnarray}
\nonumber \mathcal{S} (q) \approx \mathcal{F}_0 (q) &+& \frac{\hbar \mu \xi^2}{g} \int_0^{\mu \beta} ds \int dy \cdot y n_q \sum_{j=-\infty}^{\infty} \delta_{-j} \left(\frac{n_q}{2} + \frac{1}{2} L_{jq} \right) \delta_j \\
\label{expr_action8} &+& \frac{\hbar \mu \xi^2}{g} \int_0^{\mu \beta} ds \int dy \cdot y n_q \sum_{j=-\infty}^{\infty} \zeta_{1-j} \left(-2 L_{jq} \right) \zeta_{1j} \\
\nonumber &+&  \frac{\hbar \mu \xi^2}{g} \int_0^{\mu \beta} ds \int
dy \cdot y n_q \sum_{j=-\infty}^{\infty} \zeta_{1-j}
\left(\frac{\partial}{\partial s} + \frac{4 q j}{y^2} \right)
\delta_j  .
\end{eqnarray}

This action can be further simplified. Define
\begin{equation}
\label{def_hatLjq} \hat{L}_{jq} = \left[-\frac{1}{y} \frac{\partial}{\partial y} \left(y \frac{\partial}{\partial y}\right) + \frac{j^2}{y^2}\right] + \left(\frac{q^2}{y^2} - \frac{1-n_q}{2}\right) .
\end{equation}
Suppose  $F$ and $f$ are related by $F = \frac{f}{\sqrt{n_q}}$, then
$L_{jq}$ and $\hat{L}_{jq}$ are related by
\begin{equation}
\label{rel_Ljq} L_{jq} F = \frac{1}{\sqrt{n_q}} \hat{L}_{jq} f .
\end{equation}
Define the covariant differential operator,
\begin{equation}
\label{def_Ds} D_{s,jq} = \frac{\partial}{\partial s} - \frac{4 q j}{y^2} ,
\end{equation}
which can be seen as the time-derivative in a frame corotating with
the vortex.  With the transformation of the fluctuations,
\begin{equation}
\label{transform1} \delta_j = \frac{\hat{\delta}_j}{\sqrt{n_q}}, \zeta_{1j} = \frac{\hat{\zeta}_{1j}}{\sqrt{n_q}} ,
\end{equation}
the action (\ref{expr_action8}) is then rewritten as
\begin{eqnarray}
\label{expr_action10} \mathcal{S}(q) &\approx& \mathcal{F}_0 (q) + \frac{\hbar \mu \xi^2}{g} \int_0^{\mu \beta} ds \int dy \cdot y \sum_{j=-\infty}^{\infty}  \\
\nonumber && \left[\hat{\delta}_{-j} \frac{n_q+\hat{L}_{jq}}{2} \hat{\delta}_j - \hat{\zeta}_{1-j} (2 \hat{L}_{jq})  \hat{\zeta}_{1j} + \hat{\zeta}_{1-j} D_{s,jq} \hat{\delta}_j \right]  .
\end{eqnarray}
From (\ref{lineqnmotion1}) and (\ref{lineqnmotion2}), or from the
action (\ref{expr_action10}), the equations of motion in terms of the
new operators are
\begin{eqnarray}
\label{linequation1_new} (n_q + \hat{L}_{jq}) \hat{\delta}_j = D_{s,-jq} \hat{\zeta}_{1j} ,\\
\label{linequation2_new}  4 \hat{L}_{jq} \hat{\zeta}_{1j} = D_{s,jq} \hat{\delta}_j.
\end{eqnarray}


\section{Lifetime of the Condensate} \label{sec_lifetime}

Consider a Bose gas confined to a region of size $L$. We define as
our system a part of the Bose gas with linear size $l$ smaller than
$L$ but greater than the healing length $\xi$, i.e., $L \gg l\gg\xi$.
The Bose gas within this system is interacting with other atoms
outside, which act as a reservoir of energy, particle number and
angular momentum. Therefore the total vorticity $q$ of our system is
not conserved. The equilibrium state is described by the partition
function \cite{AltSim06_Ch9}
\begin{equation}
\label{expr_Z} \mathcal{Z} = \sum_{q=-\infty}^{\infty} \int d\delta d\zeta_1 \int D[\delta] D[\zeta_1] \exp\left(-\frac{\mathcal{S}(q)}{\hbar}\right) ,
\end{equation}
where the periodic boundary conditions \cite{NegOrl98_Ch2_2}
$\delta=\delta(0)=\delta(\mu\beta)$ and
$\zeta_1=\zeta_1(0)=\zeta_1(\mu\beta)$ have been incorporated in the
evaluation of the path integral.

Setting an upper cutoff at $y=\Lambda$, the Euclidean action (essentially the free energy divided by $k_B T$) of the system
with no vortex is obtained by putting $n_0=1$ in (\ref{def_F0})
\begin{equation}
\label{ans_F0} \mathcal{F}(0) = - \frac{ \pi \hbar \xi^2 \mu^2 \beta}{g} \int_0^{\Lambda} dy \cdot y = - \frac{ \pi \hbar \xi^2 \mu^2 \beta}{g} \frac{\Lambda^2}{2} ,
\end{equation}
and that with one vortex of vorticity $q$ is obtained after putting
the asympotic expressions (\ref{nq_largey}) and (\ref{nq_smally}) in
(\ref{def_F0}),
\begin{eqnarray}
\nonumber \mathcal{F}(q) &=& - \frac{ \pi \hbar \xi^2 \mu^2 \beta}{g} \int_0^{\Lambda} dy \cdot y n_q^2 \\
&\approx& - \frac{ \pi \hbar \xi^2 \mu^2 \beta}{g} \left(\frac{\Lambda^2}{2} - 4 q^2 \ln \Lambda\right) ,
\end{eqnarray}
as at small $y$ the integral vanishes in both cases.  As a result,
adding a vortex means adding an amount of the Euclidean action
\begin{equation}
\label{freenergy_vortex} \Delta \mathcal{F} (q) = \frac{4 \pi \hbar \xi^2 \mu^2 \beta q^2}{g} \ln \Lambda.
\end{equation}

Suppose $K_0$ is the fluctuation factor calculated from the path
integral in (\ref{expr_Z}) around the $q=0$ configuration, and $K_0
K_1$ is that around the $q=1$ configuration. If
${\kappa_{jq}^{\alpha}}^2$'s are the eigenvalues of $\hat{L}_{jq}$,
then the partition function of the $q=0$ case is given by
\cite{Fey72_Ch3}
\begin{equation}
\label{expr_K0} K_0 = \prod_{\alpha, j} \frac{1}{2 \sinh \frac{\sqrt{4 {\kappa_{j0}^{\alpha}}^2 (1+{\kappa_{j0}^{\alpha}}^2)} \mu \beta}{2}} .
\end{equation}

For $q=1$, the translation invariance of the vortices gives rise to
the existence of the zero modes \cite{Raj87_Ch10}. We know that we can
generate solutions with $\omega =0$ by simply moving the vortex
around. Since the vortex is already rotation invariant, it is enough
to consider a vortex centered at $x=R$. The displaced vortex solution
is given by $n=n_q\left(y'\right)$, $\zeta=q\varphi'$, where
\be
y'=\sqrt{y^2+R^2-2yR\cos\varphi}, \\
y'\sin\varphi'=y\sin\varphi,
\te
For small $R$ we have 
\be
y'=y-R\cos\varphi, \\ 
\varphi'=\varphi+\frac Ry\sin\varphi, 
\te
and the deviation from the
centered vortex is 
\be
\bar{\zeta}_1=\frac {qR}y\sin\varphi, \\
\bar{\delta} =-R\frac1{n_q}\frac{dn_q}{dy}\cos\varphi
\te
Then the zero-mode action $S_0$ is given by
\begin{equation}
\label{expr_S0} S_0 \approx S[\bar{\delta}, \bar{\zeta}_1] .
\end{equation}
If $\hat{\Lambda}_0$ is the operator for $q=0$ and $\hat{\Lambda}_1$
for $q=1$, then the fluctuation factor  is given by
\begin{equation}
\label{expr_KB} K_1 = \left[ \frac{\det\hat{\Lambda}_0}{\det\hat{\Lambda}_1} \right]^{\frac{1}{2}} = \sqrt{\frac{S_0}{2 \pi \hbar}}  \left[ \frac{\det\hat{\Lambda}_0}{\det'\hat{\Lambda}_1} \right]^{\frac{1}{2}} ,
\end{equation}
where the second equality  is due to the existence of a zero mode
because of the translational invariance of the vortices \cite{Col85_Ch7,
AffDeL79}, and $\det'$ is the determinant excluding the zero mode.
The ratio of the determinants is given by the Gelfand-Yaglom theorem
\cite{GelYag60, Dun08}.

Now consider the situation where more than one vortex is formed. In
the dilute gas approximation the vortices are assumed to be far apart
so adding $n$ vortices of vorticity $q=1$ increases the Euclidean action
by $n \Delta \mathcal{F} (1)$ \cite{CalCol77, Col85_Ch7}. Form a
statistical ensemble of different numbers of vortices $n$, the
partition function is given by 
\begin{eqnarray}
\nonumber \mathcal{Z} &\approx& e^{-\frac{\mathcal{F}_0(0)}{\hbar}} \sum_{n=0}^{\infty} \int d^2 y_1 \int d^2 y_2 \ldots \int d^2 y_n \cdot \frac{1}{n!} K_0 (K_1)^n e^{-n \frac{\Delta \mathcal{F}}{\hbar}} \\
\label{expr_nvortices} &=& K_0 \exp\left(-\frac{\mathcal{F}_0(0)}{\hbar} + \Lambda^2 K_1 e^{-\frac{\Delta \mathcal{F}}{\hbar}}\right)  ,
\end{eqnarray}
where the integrations over the space have an upper cutoff $\Lambda^2$,  and the factor $\frac{1}{n!}$ is due to the
indistinguishability of the vortices. The decay probability per unit
time of the configuration from $q=0$ to $1$ is \cite{CalCol77} 
\begin{equation}
\label{expr_Gamma} \Gamma = -\frac{2}{\hbar} \SIm \mathcal{F} = \frac{1}{\hbar\beta} \SIm (K_1 ) e^{-\frac{\Delta \mathcal{F}(1)}{\hbar} + 2 \ln \Lambda} .
\end{equation}
From the expression of the decay probability, the BKT transition
temperature can be read off from the exponential factor since the
formation occurs at a reasonable rate as $e^{-\frac{\Delta
\mathcal{F}}{\hbar} + 2 \ln \Lambda} \sim 1$. It is given by
\begin{equation}
\label{T_BKT} T_{BKT} \approx \frac{2 \pi \xi^2 \mu^2}{g k_B} = \frac{\pi \hbar^2 \rho_0}{2 m k_B},
\end{equation}
where the definition of healing length $\xi$ in (\ref{def_xi}) is
used and $\rho_0 = \mu / g$ \cite{GinPit58} 
is the number density of the lowest macroscopically occupied state of the homogeneous
configuration $n_0 = 1$. This agrees with the known results in the original BKT theory \cite{KosTho73}. \footnote{Another way of writing
the equation is $\rho_0 \lambda^2 = 4$ \cite{PrRuSv01},
where $\lambda = \sqrt{\frac{2\pi\hbar^2}{m k_B T}}$ is the thermal
length.} 
The correction due to non-homogeneneous configuration in the transition temperature is given in Ref. \cite{HoChWe08, HolKra08}. 

\begin{figure}[htp]
\centering
\includegraphics[scale=.8]{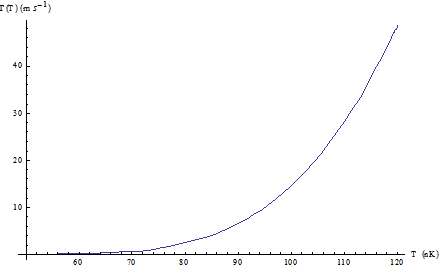}
\caption[g34] {The rate of vortex nucleation around the BKT transition temperature.} \label{g34}
\end{figure}

Because $K_1$ is given as the square root of the ratio of the determinants of two differential operators, we expect it is of order $1$. Then by dimensional analysis, $\Gamma \sim \frac{1}{\hbar\beta} \sim 13.1$ ms${}^{-1}$. 
\footnote{The cutoff is the dimensionless length of the size of the Bose gas, which is set to be $\Lambda = \sqrt{\frac{R_{\bot}}{\xi}} \sim 40$.} 
The average time of vortex formation is then of the order of $0.08$ ms. The numerical estimation of the vortex nucleation rate around the transition temperature with the estimated numerical parameters listed in Sec. \ref{sec_GPtreatment} is plotted as shown in Fig. \ref{g34}.
The vortex formation is very slow below the transition temperature but it increases drastically when the temperature increases past the critical point.

\section{Computing the imaginary part of the free energy} \label{sec_imF}

The expression for the decay rate in (\ref{expr_Gamma}) shows that under the dilute gas approximation
the stability of the canonical equilibrium hinges on whether the path
integral over fluctuations around a one-vortex configuration is
complex.

Recall that the action for a linearized fluctuation is given by
(\ref{expr_action10}), where the $\hat{L}_{jq}$ operators are defined
in (\ref{def_hatLjq}).  We perform the Gaussian  path integration
over $\zeta_1$ to obtain
\begin{eqnarray}
\nonumber \mathcal{S}(q) &\approx& \mathcal{F}_0 (q) + \frac{\hbar \mu \xi^2}{g} \int_0^{\mu \beta} ds \int dy \cdot y \sum_{j=-\infty}^{\infty} \\
\label{expr_action11} && \left[\hat{\delta}_{-j} \frac{n_q+\hat{L}_{jq}}{2} \hat{\delta}_j + \frac{1}{8} (\hat{L}_{jq}^{-1} D_{s,-jq} \hat{\delta}_{-j}) (D_{s,jq} \hat{\delta}_j) \right]  .
\end{eqnarray}
If the $\hat{L}_{jq}$ operators are positive definite, it is clear that the path of steepest descent away from the stationary point corresponds to real $\delta_j$, and the path integral is real. 

This is indeed so when we are considering fluctuations around a
homogeneous configuration, namely $q=0$, $n_q=1$. In this case, the
eigenvectors of $\hat{L}_{j0}$ are Bessel functions of order $j$. The
requirement that the Euclidean action must be finite means that we
only need to consider eigenfunctions which do not diverge at infinity
and are regular at the origin. The only Bessel functions satisfying
these conditions are of the form $J_j\left[\kappa y\right]$ corresponding
to a positive eigenvalue $\kappa^2$. Thus we conclude that the no-vortex
state is stable at zero temperature, when the no-vortex configuration
is the only finite action extremal point in the partition function.

\begin{figure}[htp]
\centering
\includegraphics[scale=.8]{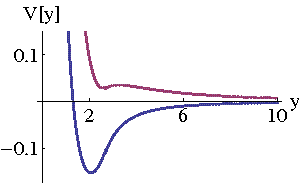}
\caption[g33] {The effective potential Eq. (\ref{pot}) for $j=1$
(upper curve) and $j=0$ (lower curve). We have used the profile in
Fig \ref{g31} to compute $n_q$. We see that for $j=1$ there can be no
negative energy state, even more so for larger values of $j$. For
$j=0$, on the other hand, the potential is negative for large enough
values of $y$. The existence of a negative energy states is shown by
Bohr-Sommerfeld condition as in (\ref{bohrsom}).} \label{g33}
\end{figure}

Let us see if this argument carries over for nonzero $q$. For
simplicity, we set $q=1$ (however, we shall leave $q$ explicit). We
seek finite action solutions to the equation

\be
\label{secular}\hat{L}_{jq}F^{\kappa}_{jq}\left(y\right)=-\kappa^2F^{\kappa}_{jq}\left(y\right) ,
\te
with real $\kappa$. The further change of variables

\be
F^{\kappa}_{jq}\left(y\right)=\frac{f^{\kappa}_{jq}\left(y\right)}{\sqrt{y}} ,
\te
reduces the left hand side to a Schrodinger operator

\be
\label{sch}\left[-\frac{d^2}{dy^2}+V_{jq}\left(y\right)\right]f^{\kappa}_{jq}\left(y\right)=-\kappa^2f^{\kappa}_{jq}\left(y\right) ,
\te
where

\be
\label{pot}V_{jq}\left(y\right)=\frac{j^2+q^2-\frac14}{y^2}-\frac12\left(1-n_q\left(y\right)\right) .
\te Therefore the question of whether the Euclidean action for
linearized fluctuations around an isolated vortex is  positive
definite becomes whether a one-dimensional particle of mass $1/2$ in
the potential (\ref{pot}) admits a negative energy state. Now, the
potential happens to be everywhere positive for all $j>0$, so we may
discard this possibility outright unless $j=0$ (see Fig. \ref{g33})

In the $j=0$ case there is a well defined potential well, and we must investigate whether it is deep enough to support a bound state. One possibility is to check the Bohr-Sommerfeld condition, namely, whether there is a value of $\kappa$ such that

\be
\label{bohrsom}\int_{y_-}^{y_+}dy\:\sqrt{-\kappa^2-V_{01}\left(y\right)}=\frac{\pi}2 ,
\te
where $y_{\pm}$ are the classical turning points, namely the roots of $\kappa^2+V_{01}\left(y\right)=0$. The answer turns out to be yes though just barely. Under the approximation given in Fig. \ref{g31} for the density profile, the Bohr-Sommerfeld condition is satisfied for $\kappa =0.024$. The turning points are located at $y_-=1.32$ and $y_+=21.2$. Bohr-Sommerfeld quantization would also predict bound excited states; however, these states fall beneath the accuracy of our approximations, and they may be considered artifacts. For example, according to Bohr-Sommerfeld quantization the first excited state appears at $\kappa =4 \times 10^{-5}$, with the outer turning point at $y=12,500$. This is beyond the intended size of the original homogeneous patch, 
because from the numerical estimation in Sec. \ref{sec_GPtreatment}, the size of the patch in the dimensionless unit is $\sqrt{\frac{R_{\bot}}{\xi}} \sim 40$, which is far less than $12,500$.
\footnote{
This series of excited states is due to the fact that the integral in (\ref{bohrsom}) diverges logarithmically when $\kappa\mapsto 0$. However, the Bohr-Sommerfeld approximation breaks down in this limit. This can be seen by approximating the density profile as $n_1=1-2/y^2$ for $y>\sqrt{2}$, $n_1=0$ otherwise. In this case the Bohr-Sommerfeld integral displays the same small-$\kappa$ behavior, but (\ref{sch}) may be solved analytically and shows no bound states. The existence of a negative energy solution to (\ref{sch}) depends critically on the effective potentiall being deeper than just $1/y^2$, and may be confirmed by independent perturbative calculations.
}

Observe that not only have we shown that the Euclidean action for
axially symmetric perturbations of the isolated vortex is not
positive definite, but we have also characterized the  eigenvector
corresponding to the direction in configuration space where it
becomes negative. Since $n_q$ does not commute with $\hat{L}_{01}$,
this eigenvector does not correspond to an actual solution of the
linearized fluctuations. However, its existence is enough to show
that the free energy acquires an imaginary part.

\section{Summary and Discussions} \label{sec_summary}

In this work we have calculated the rate of decay of an effectively homogeneous 2D Bose
gas (described by $V(\mathbf{x})=0$), in the form of $A e^{-\frac{B}{T}}$, which complies with the
well-known Arrhenius law. The prefactor $A$ is proportional to the
imaginary part of the fluctuation factor of the free energy of a
one-vortex configuration in the path integral. It is known that this
imaginary part is due to the negative eigenvalue of the fluctuation
operator belonging to the eigenvector that defines the direction the
fluctuation spontaneously grows along (in real time). The qualitative
features are like those in the decay of a metastable state due to
classical fluctuations \cite{Lan69} and barrier penetration due to
quantum fluctuations around the instanton solution of the Euclidean
action \cite{CalCol77}. We find that the imaginary part comes from
the axially-symmetric modes for nonzero vorticity configurations. As
a result, we conclude that while at $T=0$, the gas without any vortex
is stable, the canonical ensemble of different numbers of vortices of
the gas is unstable at any finite temperature.

Using the fact that at the BKT transition $A e^{-\frac{B}{T}} \sim 1$
we derived the BKT transition temperature ${T_{BKT}}$ in terms of the
number density of the homogeneous phase given in (\ref{T_BKT}). This
expression derived via thermal field theory provides a more
quantitative alternative to that originally derived from
thermodynamics considerations of the competition between the energy
and the entropy of a vortex \cite{KosTho73}. It is known that
isolated free vortices are formed in the normal phase above the BKT
temperature. Hence the decay rate calculated here is also the rate of
vortex formation. Our calculations show how it increases with
temperature. The probability of the creation of vortex pairs in a
trapped gas increases with temperature as well, 
as indicated by simulation studies with the PGPE \cite{SimBla06}." 

In the experiments,  measurements on the gas are made some time after
the confining potential is abruptly turned off. In Dalibard group's
experiment there are more isolated free vortices (measured by the
dislocation of interference pattern of two planes of gas) at higher
temperature after 20 ms time of flight (TOF) \cite{HKCBD06}. Our
results are consistent with this finding in that the rate of isolated vortex
formation increases with temperature. 
In Phillips group's experiment \cite{CRRHP09}, they observed different 
characteristics on the density profile below and above the BKT 
temperature after 10 ms TOF, which is at a rate slower than the rate of the formation
of isolated vortices.
However, since we have assumed a
homogeneous, time-independent configuration as starting point, this
should factor in the comparison of our results with experiments.
Further studies to bridge these gaps are desirable.

\section*{Acknowledgments} 

K-Y Ho thanks Prof Theodore Kirkpatrick for his interest and support, and Anand Ramanathan for useful descriptions of the Phillips group's experiment. This work is supported in part by CONICET, ANPCyT and University of Buenos Aires (Argentina), grants from NIST in the cold atom program, NSF in the ITR program and under grant No. DMR-09-01902 to the University of Maryland.

\bibliography{2DBoseGas}

\end{document}